\documentclass[12pt,pre,showpacs,preprint]{revtex4}
\usepackage{amsmath,graphicx,color,setspace,subfigure}
\usepackage{amsthm,amssymb}
\usepackage{psfrag}
\renewcommand{\eqref}[1]{(\ref{#1})}

\begin{document}

\title{Response Theory for Equilibrium and Non-Equilibrium Statistical Mechanics: Causality and Generalized Kramers-Kronig relations}

\author{Valerio Lucarini}\email{lucarini@adgb.df.unibo.it}
\affiliation{Department of Physics, University of Bologna, Bologna, Italy}
\begin{abstract}
We consider the general response theory recently proposed by Ruelle for describing the impact of small perturbations to the non-equilibrium
steady states resulting from Axiom A dynamical systems. We show that the causality of the response functions entails the possibility of writing
a set of Kramers-Kronig relations for the corresponding susceptibilities at all orders of nonlinearity. Nonetheless, only a special class of
directly observable susceptibilities obey Kramers-Kronig relations. The apparent contradiction with the principle of causality is also
clarified. Specific results are provided for the case of arbitrary order harmonic response, which allows for a very comprehensive Kramers-Kronig
analysis and the establishment of sum rules connecting the asymptotic behavior of the harmonic generation susceptibility to the short-time
response of the perturbed system. These results set in a more general theoretical framework previous findings obtained for optical Hamiltonian
systems and simple mechanical models, and shed light on the very general impact of considering the principle of causality for testing
self-consistency: the described dispersion relations constitute unavoidable benchmarks that any experimental and model generated dataset must
obey. In order to gain a more complete picture, connecting the response theory for equilibrium and non equilibrium systems, we show how to
rewrite the classical response theory by Kubo for systems close to equilibrium so that response functions formally identical to those proposed
by Ruelle, apart from the measure involved in the phase space integration, are obtained. Finally, we briefly discuss how the presented results,
taking into account the chaotic hypothesis by Gallavotti and Cohen, might have relevant implications for climate research. In particular,
whereas the fluctuation-dissipation theorem does not work for non-equilibrium systems, because of the non-equivalence between internal and
external fluctuations, Kramers-Kronig relations might be more robust tools for the definition of a self-consistent theory of climate change.
\end{abstract}
\date{\today}
\maketitle  \tableofcontents
\newpage

\section{Introduction}

The analysis of how systems respond to external perturbations to their steady state constitutes one of the crucial subjects of investigation in
the physical and mathematical sciences. In the case of physical systems near equilibrium, the powerful approach introduced by Kubo
\cite{Kubo57}, based on the generalization up to any order of nonlinearity of the formalism of the Green function, allows for expressing the
change in the statistical properties of a general observable due to the introduction of a perturbation in terms of expectation values of
suitably defined quantities evaluated at the unperturbed state \cite{Zubarev}. These results have had  huge impact on statistical mechanics and
have allowed detailed treatment of several and rather diverse processes, including, \textit{e.g.} the interaction of radiation with condensed
matter. Recently, Ruelle \cite{Rue98,Rue98b} has extended some of the investigations by Kubo to a wide class of systems \textit{far} from
equilibrium, and introduced a perturbative approach for computing the response of systems driven away by a small external forcing from their
non-equilibrium steady states. More precisely, the results by Ruelle consider perturbations to autonomous Axiom-A flows (and maps) defined in a
compact manifold, which possess a chaotic, mixing dynamics, and are associated to an invariant ergodic Sinai-Ruelle-Bown (SRB) measure
\cite{ER,Rue89}. We remind that, the - mathematically speaking, special - case of Axiom A systems amounts to being of general physical interest,
if one accepts the chaotic hypothesis by Gallavotti and Cohen \cite{Gallavotti95b,Gallavotti96} which states that, for the purpose of computing
macroscopic quantities, many-particle systems behave \emph{as though} they were dynamical systems with transitive Axiom-A global attractors.
Ruelle shows that, in analogy to what found by Kubo, at all orders of perturbative expansion, the effect of the forcing on the expectation value
of a general observable can be expressed in terms of averages of quantities performed at the non-equilibrium steady state, \textit{i.e.}
obtained by integrating over the unperturbed SRB measure. Moreover, in the case of linear response, it is shown that it is possible to define
formally a susceptibility function, obtained as the Fourier Transform of the linear Green function of the system, and to prove that such
susceptibility, basically as a result of the causality principle, obeys Kramers-Kronig (K-K) relations \cite{Nus72,Peip99}, just as in Kubo
framework. The K-K relations say that the the real and imaginary part of the linear susceptibility are fundamentally connected, each one being
the Hilbert transform of the other one. Hence, these integral properties provide unavoidable constraints for checking the self-consistency of
experimental or model-generated data. Furthermore, by applying the K-K relations, it is possible to perform the so-called inversion of data,
\textit{i.e.} to acquire knowledge on the real part by measurements of the imaginary part over the whole spectrum or vice versa.

Nevertheless, in spite of such important formal analogies, it is important to stress some qualitative differences between equilibrium and
non-equilibrium systems in the physical meaning of the linear response function. Whereas in systems close to equilibrium there is basically
equivalence between the natural fluctuations and the linear response to external perturbations, as clarified by the fluctuation-dissipation
theorem \cite{Weber56,Kubo66}, in the considered non equilibrium systems such symmetry in broken, the mathematical reason being that the SRB
measure is smooth only along the unstable manifold. A more geometrical view of this fact is that, whereas natural fluctuations of the system are
restricted to the unstable manifold, because, by definition, asymptotically there is no dynamics along the stable manifold, external forcings
will cause almost always motions having components - of exponentially decaying amplitude - out of the unstable manifold \cite{Rue98,Rue98b}. For
a discussion of this point, see also \cite{Cessac}. It should also be noted that the non-equivalence of forced and free fluctuations in chaotic
systems was already pointed out and tackled in heuristic terms in the late '70s by Lorenz \cite{Lorenz79} when considering the atmospheric
system.

Whereas K-K relations for linear processes, thanks to their generality, have become a basic textbook subject and standard tool in many different
fields, such as acoustics, signal processing, optics, statistical mechanics, condensed matter physics, material science, relatively little
attention has been paid to theoretical and experimental investigation of K-K relations and sum rules of the nonlinear susceptibilities, in spite
of the ever increasing scientific and technological relevance of nonlinear physical processes. Recently, several theoretical and experimental
results in this direction have been formulated in the context of analyzing nonlinear processes of interaction of radiation with matter
\cite{Luc03,Luc05}.

The main goal of this paper is to analyze the formal properties on the $n^{th}$ order perturbative response of Axiom A, non equilibrium steady
state systems to external forcings. In particular, we develop a theory of generalized K-K relations that extend, on one side, the results on the
linear case given by Ruelle \cite{Rue98} for this class of systems, and on the other side, what obtained for nonlinear processes in electronic
systems close to equilibrium \cite{Luc03,Luc05} and in simple yet prototypical mechanical systems \cite{Luc98}. Special attention is paid to the
case of nonlinear susceptibilities describing processes responsible for harmonic generation, whose properties are such that a rather extensive
set of important properties - including sum rules - can be deduced. We stress that also in the nonlinear case K-K relations constitute
unavoidable benchmarks that any experimental and model generated dataset must obey. K-K relations may prove, as discussed later, useful tools
for defining a theory of climate change, because they apply also for systems where the fluctuation-dissipation theorem is not verified.

Our paper is structured as follows. In Sec. \ref{se2}, we introduce the properties of the general $n^{th}$ susceptibility, resulting as Fourier
transform of the $n^{th}$ perturbative order response function of the system. In Sec. \ref{se3}, we present an extension of the theory of
nonlinear K-K relations to the dynamical systems considered by Ruelle, showing which class of nonlinear susceptibilities obey K-K relations and
deducing rigorous results in the case of harmonic generation processes. In Sec. \ref{se4}, we discuss our results, present our conclusions and
perspectives for future investigations. Two appendices are also included. In App. \ref{kubo} we show how the Kubo theory can be formally
reconciled with the results by Ruelle, so that the results presented in this work can be applied also for general equilibrium systems. A
discussion of the relevance for climate studies of the response theory for Axiom-A systems and of the specific results described in this study
is given in App. \ref{climate}.

\section{Linear and nonlinear response of perturbed non-equilibrium steady states}\label{se2}
We consider an autonomous Axiom-A flow $\dot{x}=F(x)$ defined in a compact manifold, such that $x(t)=f^t x$, with $x=x(0)$. The flow is assumed
to possess a chaotic, mixing dynamic, and to be associated to an invariant ergodic SRB measure $\rho_{SRB}(\textrm{d}x)$, such that for any
measurable observable $\Phi(x)$ the ensemble average is equal to the time average:
\begin{equation}
\langle \Phi \rangle_0 =\int \rho_{SRB}(\textrm{d}x) \Phi(x) = \lim_{T\rightarrow\infty}\frac{1}{T} \int\limits_{0}^{T }\textrm{d}t
\Phi(f^{t}x)= \lim_{T\rightarrow\infty}\int \textrm{d}x \Phi(f^{T}x)
\end{equation}
for almost every initial condition $x$ according to the Lebesgue measure d$x$; the last equality holds for the special case of mixing dynamics.
The SRB measure is usually singular, but smooth along the directions of the unstable manifold \cite{ER,Rue89,You02}. Ruelle has shown that for
this specific class of dynamical systems (as well as for the corresponding discrete-time diffeomorphisms) it is possible to differentiate the
SRB states \cite{Rue97,Rue03} when the flow is perturbed by an infinitesimal vector field in the following way:
\begin{equation}
\dot{x}=F(x)+e(t)X(x).
\end{equation}
Is is then possible to express the perturbed expectation value of $\Phi(x)$ in terms of a perturbation series:
\begin{equation}
\langle \Phi \rangle (t) = \langle \Phi \rangle_0 + \sum_{n=1}^{\infty} \langle \Phi \rangle^{(n)}(t)
\end{equation}
where, proposing a slight generalization of the formula proposed by Ruelle \cite{Rue98b}, which considered purely periodic perturbations, the
$n^{th}$ term can be expressed as a \textit{n}-uple convolution integral of the $n^{th}$ order Green function with \textit{n} terms each
representing the suitably delayed time modulation of the perturbative vector field:
\begin{equation}
\langle \Phi \rangle^{(n)}(t) =\int\limits_{ - \infty }^{ \infty }\int\limits_{ - \infty }^{ \infty }\ldots \int\limits_{ - \infty }^{ \infty
}\textrm{d}\sigma_1\textrm{d}\sigma_2\ldots \textrm{d}\sigma_n G^{(n)}(\sigma_1,\ldots,\sigma_n)e(t-\sigma_1)e(t-\sigma_2)\ldots e(t-\sigma_n).
\label{phin}
\end{equation} The $n^{th}$ order
Green function is causal, \textit{i.e.} its value is zero if any of the argument is non positive, and can be expressed as time dependent
expectation value of an observable evaluated over the unperturbed SRB measure:
\begin{eqnarray}
G^{(n)}(\sigma_1,\ldots,\sigma_n)=\int \rho_{SRB}(\textrm{d}x)&&\Theta(\sigma_1)\Theta(\sigma_2-\sigma_1)\ldots\Theta(\sigma_n-\sigma_{n-1})\times \nonumber\\
&&\times\Lambda\Pi(\sigma_n-\sigma_{n-1})\ldots\Lambda\Pi(\sigma_2-\sigma_1)\Lambda\Pi(\sigma_1)\Phi(x), \label{greenrue}
\end{eqnarray}
where $\Theta$ is the usual Heaviside function, $\Lambda(\bullet)=X(x)\nabla(\bullet)$ describes the effect of the perturbative vector field,
and $\Pi$ induces the time evolution along the unperturbed vector field so that $\Pi(\tau)A(x)=A(x(\tau))$ for any observable $A$. The $n=1$
term describes the linear response of the system to the perturbation field \cite{Rue98}, and, thanks to the superposition principle, can be
derived also by using the method of impulse perturbation \cite{Cessac}. In App. \ref{kubo} we show that it is possible to rephrase the Kubo
response theory \cite{Kubo57} in such a way to obtain a formula that perfectly matches the formula presented in Eq. \eqref{greenrue}, provided
that the equilibrium canonical distribution is used instead of the SRB measure $\rho_{SRB}(\textrm{d}x)$.

\subsection{Response of the system in the frequency domain}
If we compute the Fourier transform of the $n^{th}$ order perturbation to the expectation value $\langle \Phi \rangle^n(t)$ defined in Eq.
\eqref{phin} we obtain:
\begin{equation}\label{PolNL}
\langle \Phi \rangle^{(n)} ( \omega ) = \int\limits_{ - \infty }^{ \infty }\ldots \int\limits_{ - \infty }^{ \infty }\textrm{d}\omega _{1 }
\ldots \textrm{d}\omega _{n} \chi^{( n )} \left( {\omega _{1 } ,\ldots,\omega _{n } }\right)  e( {\omega _{1 } } )\ldots e( {\omega _{n} } )
\times\delta\left( {\omega - \sum\limits_{l=1}^{n}{\omega_l} } \right),
\end{equation}
where the Dirac $\delta$ guarantees that the sum of the arguments of the Fourier transforms of the time modulation functions equals the argument
of the Fourier transform of $\langle \Phi \rangle^n(t)$, whereas the susceptibility function is defined as
\begin{equation}
\label{eq4} \chi^{( n )} \left({\omega _{1 } ,\ldots,\omega _{n } } \right) = \int\limits_{ - \infty }^{ \infty }  \ldots \int\limits_{ - \infty
}^{ \infty }\textrm{d}t_{1 } \ldots \textrm{d}t_{n } G^{ (n) } \left( {t_{1 }, \ldots , t_{n } }
\right)\exp\left[\textrm{i}\sum\limits_{j=1}^{n}{\omega_j t_j} \right].
\end{equation}
These operations make sense if the Green function is integrable or at least, in a weaker, distributional sense, if it not exponentially
increasing. In the linear case, Ruelle \cite{Rue98,Rue98b} has shown that integrability is ensured by proving that both the contributions
associated to terms resulting from projections of the perturbative vector field on the unstable and stable manifolds converge, because of the
distinct processes of mixing and of exponential contraction, respectively. In the nonlinear $n>1$ case, we can heuristically use the same
arguments - as well as taking into account that in the classical equilibrium case \cite{Kubo57} the higher order correlations are typically much
weaker and with faster decrease - to exclude the possibility that the operation presented in Eq. \eqref{eq4} is meaningless.

Assuming that, without serious loss of generality, the function $e(t)$ can be expressed as:
\begin{equation}
\label{et} e(t)= \sum\limits_{k=1}^{m} e_{\omega_j} \exp[-\textrm{i}\omega_j t] + e_{-\omega_j} \exp[\textrm{i}\omega_j t]
\end{equation}
with $e_{\omega_j}=\left[e_{-\omega_j}\right]^{\ast}$, we derive that each frequency component in Eq. \ref{PolNL} can be written as:
\begin{equation}
\label{eq6} \langle \Phi \rangle^{(n)} ( \omega )  = \sum \limits_{\{\omega_\Sigma\}}\overline{\langle \Phi \rangle}^{(n)}\left( \omega_\Sigma
\right)\delta\left(\omega-\omega_\Sigma\right),
\end{equation}
where we are summing over all the possible distinct values $\{\omega_\Sigma\}$ of the possible sums of $n$ among the $2m$ frequencies in the
spectrum of $e(t)$, which basically formalizes the process of \textit{frequency mixing}. Of course, in the linear $n=1$ case, no mixing occurs
and outputs can be observed at the same frequencies as the input. In general, each term $\overline{\langle \Phi \rangle}^{(n)}\left(
\omega_\Sigma \right)$ is given by the following sum:
\begin{equation}
\overline {\langle \Phi \rangle }^{(n)}\left( \omega_\Sigma \right)=\sum\limits_{\sum\omega_{k_j}=\omega_\Sigma}\chi^{( n )} \left({\omega _{k_1
} ,\ldots,\omega _{k_n }}\right)e_{w_{k_1}}\ldots e_{w_{k_n}}, \label{respofreq}
\end{equation}
where the sum of the arguments of all the contributing susceptibility functions is $\omega_\Sigma$. Note that, from an experimental point of
view, we can measure $\overline{\langle \Phi \rangle}^{(n)}\left( \omega_\Sigma \right)$ by analyzing in the frequency domain the perturbed
output of the system, whereas disentangling the various terms contributing to the summation in Eq. \eqref{respofreq} is rather hard. Again, this
problem is not present in the linear case.

\section{Generalized Kramers-Kronig relations}
\label{se3}
\subsection{Basic Results}
Once we are granted that at every order \textit{n} the response on the system $\langle \Phi \rangle^{(n)}(t)$ is written as a convolution
integral having as Kernel a causal Green function $G^{(n)}(\sigma_1,\ldots,\sigma_n)$, and assuming that the suitable integrability conditions
are obeyed, we are in the conditions of writing generalized dispersion relations for the $n^{th}$ order susceptibility presented in Eq.
\eqref{eq4}, along the lines of what developed in the context of optics in \cite{Luc03,Luc05}. Therefore, we can apply Titchmarsch's theorem
\cite{Nus72,Peip99,Luc03,Luc05} separately to each variable of $G^{(n)}(\sigma_1,\ldots,\sigma_n)$ and deduce that $\chi^{( n )} \left({\omega
_{1 } ,\ldots,\omega _{n } } \right)$ is holomorphic in the upper complex plane of each variable $\omega_i$, $1\leq i\leq n$. If we consider the
first argument $\omega_{1}$ of the nonlinear susceptibility function (\ref{eq4}), the following dispersion relation holds
\begin{equation}\label{KKnl}
 {\rm{P}} \int\limits_{-\infty}^{\infty}\textrm{d}\omega'_{1}  \frac{\chi^{( n) } \left(\omega'_{1} ,\ldots ,\omega'_{n}\right)
}{\omega'_{1}-\omega_{1}}=\textrm{i}\pi \chi\left(\omega_{1} ,\ldots ,\omega'_{n}\right),
\end{equation}
where P indicates that integration is performed in principal part. Repeating the same procedure for all the remaining $n-1$ frequency variables,
we obtain
\begin{equation}\label{KKnl2}
 {\rm{P}} \int\limits_{-\infty}^{\infty}\ldots \int\limits_{-\infty}^{\infty}\textrm{d}\omega'_{1}\ldots\textrm{d}\omega'_{n}  \frac{\chi^{( n) } \left(\omega'_{1} ,\ldots ,\omega'_{n}\right)
}{(\omega'_{1}-\omega_{1})\ldots (\omega'_{n}-\omega_{n})}=\left(\textrm{i}\pi\right)^n \chi\left(\omega_{1} ,\ldots ,\omega_{n}\right),
\end{equation}
which extends to all orders the linear K-K relations already described by Ruelle \cite{Rue98}. K-K relations constitute self-consistency
constraints that must be obeyed and allow to reconstruct the real part of the response from the imaginary part, or vice-versa. The principle of
causality of the response function is reflected mathematically in the validity of the K-K relations presented in Eq. \eqref{KKnl2}. Note that,
as discussed by Peiponen \cite{Pei87,Pei88}, all the functions $\left[\chi^{( n) } \left(\omega_{1} ,\ldots ,\omega_{n}\right)\right]^m$, with
$m\geq1$ obey the very same dispersion relations as that written for $m=1$ in Eq. \eqref{KKnl2}. This implies that the generality of these
dispersion relation goes beyond not only the distinction between classical and quantum equilibrium system, as discussed in \cite{Luc03,Luc05},
but also beyond the distinction between equilibrium and non-equilibrium systems, at least when we consider the Axiom A case or adopt the chaotic
hypothesis for many particles systems.

The dispersion relations \eqref{KKnl} and \eqref{KKnl2} may be thought of being of doubtful interest from an experimental point of view, since
on one side we basically can have access to quantities like $\overline{\langle \Phi \rangle}^{(n)}(\omega_\Sigma)$, which results from a linear
combination of, in general, more than one different susceptibility functions (in the sense that they are evaluated at different values of their
arguments). Moreover, most of the physically relevant nonlinear phenomena are described by nonlinear susceptibilities where all or part of the
frequency variables are mutually dependent, such in the later described case of $n^{th}$ order harmonic generation at frequency $n\omega$ in the
presence of a monochromatic modulation function $e(t)=\exp\left[-\textrm{i}\omega_0 t\right]+\exp\left[\textrm{i}\omega_0 t\right]$ of frequency
$\omega_0$. We may, therefore, understand that a more flexible theory is needed in order to provide the effectively relevant dispersion
relations for nonlinear phenomena.

\subsection{A New Definition of Dispersion Relations}

We take the following point of view. When considering the $n^{th}$ order nonlinear process, a meaningful dispersion relation involves a line
integral in the space of the frequency variables, which entails the choice of a one-dimensional space embedded in a $n$-dimensional space. This
corresponds to the realistic experimental setting where only the frequency of one of the monochromatic fields described in Eq. \eqref{et} is
changed. Since in the nonlinear setting we have frequency mixing, changing the frequency of one of the components of the forcing will change
differently each of the terms $\overline {\langle \Phi \rangle }^{(n)}\left( \omega_\Sigma \right)$, depending on whether none, one or more than
one arguments of the contributing nonlinear susceptibility functions (see Eq. \eqref{respofreq}) are varied. The choice of the parameterization
then selects different susceptibilities and so refers to different nonlinear processes. Each component $j$ of the straight line in
$\mathbb{R}^{n}$ can be parameterized as follows:
\begin{equation}\label{scandolo}
\omega_j=v_js+w_j, 1 \leq j\leq n,
\end{equation}
where the parameter $s \in  \mathbb{R}$, the vector $\vec{v}\in \mathbb{R}^{n}$ of its coefficients describes the direction of the straight
line, and the vector $\vec{w}\in \mathbb{R}^{n}$ determines $\vec{\omega}(0)$. Since we know that $\chi^{( n) } \left( \omega_{1},\ldots
,\omega_{n}\right)$ is holomorphic in the upper complex plane of each variable $\omega_i$, $1\leq i\leq n$, we have that the extension for
complex values of $s$ of the function:
\begin{equation}\label{scandolo1}
\begin{split}
\chi^{\left(n\right) } \left( s\right)&=\chi^{\left( n\right) }\left(v _{{1} }s+w_1,\ldots
,v _{{n} }s+w_n  \right)\\
&=\int\limits_{-\infty }^{\infty }\ldots \int\limits_{-\infty }^{\infty }\textrm{d}t_{1} \ldots \textrm{d}t_{n} G^{\left( n\right) } \left(
t_{1} ,\ldots ,t_{n} \right) \exp\left[\textrm{i}s\sum\limits_{j=1}^{n}v_{{j} } t_{j}+\textrm{i}\sum\limits_{j=1}^{n}w_{{j} } t_{j}\right]
\end{split}
\end{equation}
is holomorphic in the upper complex $s$ plane if all the components of vector $\vec{v}$ are non-negative. This construction has been first
proposed in the context of nonlinear optics in \cite{Bas91}. Hence, by applying the Titchmarch theorem, we deduce that for all $m\geq1$ the
following integral relation holds for the susceptibility defined in Eq. \eqref{scandolo1}:
\begin{equation}\label{KKnl3a}
\textrm{i}\pi \left[\chi^{( n) } (s)\right]^m={\rm{P}}\int\limits_{-\infty}^{\infty} \frac{\left[\chi^{( n) } ( s')\right]^m
}{s'-s}\textrm{d}s',
\end{equation}
which, when the real and imaginary part of the nonlinear susceptibility are considered, results into:
\begin{equation}\label{KKnl3b}
{\rm{Re}}\left\{\left[\chi^{( n) } (s)\right]^m\right\}=\frac{1}{\pi}{\rm{P}}\int\limits_{-\infty}^{\infty} \frac{{\rm{Im}}\left\{\left[\chi^{(
n) } ( s')\right]^m\right\} }{s'-s}\textrm{d}s',
\end{equation}
\begin{equation}
\label{KKnl3c}{\rm{Im}}\left\{\left[\chi^{(n) } (s)\right]^m\right\}=-\frac{1}{\pi}{\rm{P}}\int\limits_{-\infty}^{\infty}
\frac{{\rm{Re}}\left\{\left[\chi^{( n) } ( s')\right]^m\right\} }{s'-s}\textrm{d}s'.
\end{equation}
The condition on the sign of the directional vectors of the straight line in $\mathbb{R}^{n}$ implies that only one particular class of
nonlinear susceptibilities possess the holomorphic properties required to obey the dispersion relations (\ref{KKnl3b}). Hence, causality is not
a sufficient condition for the existence of K-K relations between the real and imaginary part of a general nonlinear susceptibility function,
\textit{if its arguments are mutually dependent}. We stress that, instead, causality implies that Eq. \eqref{KKnl2} holds.
\subsection{Harmonic Generation}
In order to clarify the results presented in the previous sections, and show how they can be used for analyzing actual data, we concentrate on
the simplified setting of a single monochromatic perturbation field such that $e(t)=\exp[-\textrm{i}\omega_0 t]+\exp[\textrm{i}\omega_0 t]$. In
this case, at each order $n$, $\omega_\Sigma =\pm (2j+1)$, with $\omega_0, j=0,\ldots,(n-1)/2$ if $n$ is odd and $\omega_\Sigma =\pm
2j\omega_0$, with $j=0,\ldots,n/2$ if $n$ is even. Note that for even orders there is always a static response, which, in the optical
literature, is known as optical rectification \cite{Luc05}. If we focus, \textit{e.g.}, on the third order of perturbation and consider only the
positive frequencies, we have that the observable signal at $\omega_0$, constituting the first correction to the linear response is:
\begin{equation}\label{kerr}
\overline{\langle \phi
\rangle}^{(3)}(\omega_0)=\chi^{(3)}(-\omega_0,\omega_0,\omega_0)+\chi^{(3)}(\omega_0,-\omega_0,\omega_0)+\chi^{(3)}(\omega_0,\omega_0,-\omega_0);
\end{equation}
whereas the observable signal, responsible for the third harmonic generation is:
\begin{equation}\label{harmo} \overline{\langle \phi
\rangle}^{(3)}(3\omega_0)=\chi^{(3)}(\omega_0,\omega_0,\omega_0);
\end{equation}
If we change the frequency $\omega_0$ of the perturbation field and study how the output varies, it is clear that for all the three terms
comparing on the right hand side of Eq. \eqref{kerr} the vector $\vec{v}$ of the s-parameterization proposed in Eq. \eqref{scandolo} has one
negative component, whereas $\vec{v}=(1,1,1)$ for the only term responsible for harmonic generation in Eq. \eqref{harmo}. This implies that,
when analyzing the first nonlinear correction to the linear response at frequency $\omega_0$, we cannot expect that K-K relations apply, since
poles in the upper complex plane of the $s=\omega_0$ may well be present \cite{Pei04}. In this case, different signal processing techniques,
such as the Maximum Entropy Method, have to be adopted \cite{Luc05}. Therefore, the condition on the sign of $\vec{v}$ is in this case useful
for giving a negative statement, \textit{i.e. determining when K-K cannot be applied}. On the contrary, we are granted that the susceptibility
describing the third harmonic nonlinear response obeys K-K relation. It is clear that the same applies at all orders $n$, and also it can be
easily shown that the only contribution to the observable $\overline{\langle \phi \rangle}^{(n)}(n\omega_0)$ is
$\chi^{(n)}(\omega_0,\ldots,\omega_0)$.

At every order $n$, we have that $\chi^{( n )} \left({ - \omega _{1 } ,\ldots, - \omega _{n } } \right) = \left\{ {\chi^{( n )} \left(\omega_{1
} ,\ldots,\omega _{n }
 \right)} \right\}^\ast$ ($\left\{Z\right\}^\ast$ indicating the complex conjugate of $Z$), because $\langle \Phi \rangle ^{(n)}(t)$ and $e(t)$ are real. It is is easy to show that the following relation holds for all values of $m\geq1$:
\begin{equation}
\label{eq5b} \left[\chi^{( n )} \left({ - \omega _{1 } ,\ldots, - \omega _{n } } \right)\right]^m = \left\{\left[ {\chi^{( n )} \left(\omega_{1
} ,\ldots,\omega _{n }
 \right)} \right]^m\right\}^\ast,
\end{equation}
We then derive that at all orders $n\geq1$ and for all $m\geq1$:
\begin{equation}\label{harmo1}
-\frac{\pi}{2} {\rm{Im}}\left\{\left[\chi^{( n) } \left(\omega_0,\ldots ,\omega_0\right)\right]^m\right\}= \omega_{0}{\rm{P}}
\int\limits_{0}^{\infty} \textrm{d}\omega'_{0} \frac{{\rm{Re}}\left\{\left[\chi^{( n) } \left(\omega'_{0} ,\ldots
,\omega'_{0}\right)\right]^m\right\} }{({\omega'_{0}}^2-\omega_{0}^2)}
\end{equation}
\begin{equation}\label{harmo2}
\frac{\pi}{2} {\rm{Re}}\left\{\left[\chi^{( n) } \left( \omega_{0},\ldots ,\omega_{0}\right)\right]^m\right\} =  {\rm{P}}
\int\limits_{0}^{\infty}\textrm{d}\omega'_{0} \frac{\omega'_{0}{\rm{Im}}\left\{\left[\chi^{( n) } \left(\omega'_{0} ,\ldots
,\omega'_{0}\right)\right]^m\right\} }{({\omega'_{0}}^2-\omega_{0}^2)}
\end{equation}
which, albeit in a different perspective from what shown in Eq. \eqref{KKnl2}, generalize the linear K-K at all orders. Note that, if we
consider $\lim_{\omega_0\rightarrow 0}$ of Eq. \eqref{harmo2} in the linear case and assume that the limits converge, we obtain the following
expression for the linear static response of the system:
\begin{equation}\label{sr2}
\rm{Re}\left\{\left[\chi^{(1)}\left( 0\right)\right]^m\right\} = \left[\rm{Re}\left\{\chi^{(1)}\left( 0\right)\right\}\right]^m=\frac{2}{\pi}
{\rm{P}} \int\limits_{0}^{\infty}\textrm{d}\omega'_{0} \frac{{\rm{Im}}\left\{\left[\chi^{( 1) } \left(\omega'_{0}\right)\right]^m\right\}
}{{\omega'_{0}}};
\end{equation}
the finiteness of the integral is consistent with the fact that, by symmetry, ${\rm{Im}}\left\{\left[\chi^{( 1) }
\left(\omega_{0}=0\right)\right]^m\right\} =0$, which must be obeyed for all values of $m\geq1$. Note that a detailed verification of linear K-K
has been performed in the case of Lorenz system \cite{Reick02}. It is somewhat surprising to observe how the qualitative features of the
detected (and reconstructed) susceptibility are similar to what results from a simple oscillator model: the imaginary part has a strong peak for
a resonance of system (even if in this case there is no deterministic \textit{natural frequency} for the system), which matches the dispersive
structure found for the real part of the susceptibility. Another minor spectral feature is observed, and again, following the spectroscopic
paradigm, a peak in the imaginary part is associated to a dispersive structure in the real part. Note also that the Lorenz system is non-Axiom
A, which suggests that a wide range of applicability for these relations is still to be explored. Note that, even if several monochromatic
forcings are present, Eqs. \eqref{harmo1}-\eqref{harmo2} still apply, since no other frequency components are involved.

If we plug $\vec{v}=(1,\ldots,1)$ and $\vec{w}=(0,\ldots,0)$, and redefine $s=\omega_0$ in Eq. \eqref{scandolo1}, and consider the basic
properties of the Fourier Transform, we have that the short time behavior of the $n^{th}$ order Green function determines the asymptotic
behavior of the $n^{th}$ order harmonic susceptibility at frequency $n\omega_0$. We perform the following variable change
\begin{equation}
t_j=\sum\limits_{k=1}^{j}\tau_k,
\end{equation}
assume that $G^{(n)}(t_1(\tau_1),\ldots,t_n(\tau_1,\ldots\tau_n))$ be smooth for all its arguments $\{\tau_j\}$ in 0, and let $\beta$ be the
smallest sum of exponents of $(\tau_1,\ldots \tau_n)$ such that there is a non-vanishing monomial $M_\beta(\tau_1,\ldots,\tau_n)$ in the Taylor
expansion $G^{(n)}(t_1(\tau_1),\ldots,t_n(\tau_1,\ldots\tau_n))$. We then have that the following limit exists and is finite
\cite{Bas00,Luc03,Luc05}:
\begin{equation}\label{asy}
\lim_{\omega_0\rightarrow \infty}\omega_0^{\beta+n}\chi^{(n)}(\omega_0,\ldots,\omega_0)=\alpha\in\mathbb{R}\setminus\{0\},
\end{equation}
which implies that the asymptotic behavior of $\chi^{(n)}(\omega_0,\ldots,\omega_0)$ is at least as fast as $\omega_0^{-n}$. Moreover, since
$\textrm{Re}\left\{\chi^{(n)}(\omega_0,\ldots,\omega_0)\right\}$ is an even function of $\omega_0$ and, from Eq. \eqref{asy}, determines the
asymptotic behaviour ($\textrm{Im}\left\{\chi^{(n)}(\omega_0,\ldots,\omega_0)\right\}$ has a faster asymptotic decrease), we derive that
$\beta+n$ must be even, so that $\beta+n=2\gamma$. Therefore, dispersion theory provides us with indirect information also about the short time
behavior of the Green function. Furthermore, the knowledge of the asymptotic behavior allows a further generalization of what presented in
\eqref{harmo1}-\eqref{harmo2}. In fact, we have that all the (independent) functions $\omega^{2p}\chi^{(n)}(\omega_0,\ldots,\omega_0)$,
$p=0,\ldots,\gamma-1$ are holomorphic in the upper complex plane of $\omega_0$ and obey suitable integrability conditions allowing for writing
the following set of generalized K-K relations:
\begin{equation}\label{harmo3}
-\frac{\pi}{2} {\omega_{0}}^{2p+1}{\rm{Im}}\left\{\left[\chi^{( n) } \left(\omega_0,\ldots ,\omega_0\right)\right]^m\right\}={\rm{P}}
\int\limits_{0}^{\infty} \textrm{d}\omega'_{0} \frac{{{\omega'_0}^{2p}\rm{Re}}\left\{\left[\chi^{( n) } \left(\omega'_{0} ,\ldots
,\omega{'}_{0}\right)\right]^m\right\} }{(({\omega'_{0}}^2-\omega_{0}^2)},
\end{equation}
\begin{equation}\label{harmo4}
\frac{\pi}{2} {\omega_{0}}^{2p}{\rm{Re}}\left\{\left[\chi^{( n) } \left( \omega_{0},\ldots ,\omega_{0}\right)\right]^m\right\} =  {\rm{P}}
\int\limits_{0}^{\infty}\textrm{d}\omega'_{0} \frac{{\omega'_0}^{2p+1}{\rm{Im}}\left\{\left[\chi^{( n) } \left(\omega'_{0} ,\ldots
,\omega{'}_{0}\right)\right]^m\right\} }{({\omega'_{0}}^2-\omega_{0}^2)}.
\end{equation}
with $p=0,\ldots,m\gamma-1$. Comparing the asymptotic behavior given in  Eq. \eqref{asy} with those obtained by applying the superconvergence
theorem \cite{frye63} to the general K-K relations \eqref{harmo3}-\eqref{harmo4}, we derive the following set of general sum rules
\begin{equation} \label{SRH1}
\int\limits_0^\infty {{\omega_0 }'^{2p }} {\rm{Re}} \left\{\left[ \chi^{\left( n \right)} \left( {\omega'_0 },\ldots ,{\omega'_0 } \right)
\right]^m\right\}\textrm{d}{\omega }' = 0, \hspace{4mm} 0 \leq p \leq m\gamma-1,
\end{equation}
\begin{equation} \label{SRH2}
\int\limits_0^\infty {{\omega_0 }'^{2p+1 }} {\rm{Im}} \left\{ \left[\chi^{\left( n \right)} \left( {\omega'_0 },\ldots ,{\omega'_0 } \right)
\right]^m\right\}\textrm{d}{\omega }' = 0, \hspace{4mm} 0 \leq p \leq m\gamma-2,
\end{equation}
\begin{equation} \label{SRH3}
\int\limits_0^\infty {{\omega_0 }'^{2p+1 }} {\rm{Im}} \left\{\left[ \chi^{\left( n \right)} \left( {\omega'_0 },\ldots ,{\omega'_0 } \right)
\right]^m\right\}\textrm{d}{\omega }' = -\alpha^m \frac{\pi}{2}, \hspace{4mm} p = m\gamma-1.
\end{equation}
All the moments of the  $n^{th}$ order harmonic generation susceptibility vanish except that of order $2\gamma-1$ of the imaginary part. This
latter sum rule creates a conceptual bridge between the measurements of the imaginary part of the susceptibility under examination throughout
the spectrum to the short term behavior of the $n^{th}$ Green function. These results hold for all values of $m\geq1$. The generalized K-K
relations and sum rules here presented constitute a rather extensive set of stringent integral constraints that must be obeyed by experimental
data and model simulations. These results generalize what obtained for general optical systems near equilibrium \cite{Bas00} and for simple
mechanical systems \cite{Luc98}. Note that both the generalized K-K relations \eqref{harmo3}-\eqref{harmo4} and the sum rules
\eqref{SRH1}-\eqref{SRH3} have been verified in detail on experimental data in the case of optical processes near equilibrium \cite{Luc03b}.

\section{Summary and Conclusions}
\label{se4}

In this paper we have considered the general response function $G^{(n)}(t_1,\ldots , t_n)$ recently proposed by Ruelle \cite{Rue98,Rue98b} for
describing the impact of small time-dependent forcings to the non-equilibrium steady states resulting from Axiom A dynamical systems, which,
when taking into account the chaotic hypothesis by Gallavotti and Cohen \cite{Gallavotti95b,Gallavotti96}, are of general physical interest. At
all orders of perturbative expansion, the effect of the forcing on the expectation value of a general observable can be expressed in terms of
means of quantities performed at non-equilibrium steady state.

Since, at every order of perturbation, the response function is causal, it is possible to write a set of Kramers-Kronig relations for the
corresponding susceptibility, defined as the multivariable Fourier Transform of the response function $\chi^{(n)}(\omega_1,\ldots,\omega_n)$.
These dispersion relations are of little applicability because they cannot be used to effectively analyze the output signal, which is the change
in the expectation observable of the considered observable.

In practice, it is interesting to consider the case of one or more monochromatic forcings and to be in the condition of analyzing what happens
when the frequency of one of them is changed. Since in the nonlinear setting of order $n$ we have frequency mixing, such frequency tuning will
affect differently the various frequency components of the observed output signal, depending on whether none, one or more than one arguments of
the nonlinear susceptibility functions responsible for the observed frequency components of the output are varied. Therefore, following this
approach, the dispersion relation becomes a parameterized line integral in the $n$-dimensional space of frequency variables. K-K relations apply
only for special forms of parameterizations, which correspond to a specific family of susceptibility functions. These results are
system-independent and derive strictly from complex analysis.

Among the phenomena which can be treated using the K-K formalism, we concentrate on the $n^{th}$ order process by which the system responds at
frequency $n\omega_0$ when forced by a monochromatic vectorial field with angular frequency $\omega_0$. Such a process is described by the
harmonic generation susceptibility $\chi^{(n)}(\omega_0,\ldots,\omega_0)$, which is holomorphic in the upper complex $\omega_0$ plane and obeys
K-K relations. For any given system, the asymptotic behavior for large frequencies is shown to depend on the short-time response and to be of
the form $\omega_0^{-2\gamma}$. It is then proved that all functions $\omega^{2p}\chi^{(n)}(\omega_0,\ldots,\omega_0)$ with
$p=0,\ldots,\gamma-1$ obey K-K relations, so that more stringent , generalized constraints are established. Furthermore, using symmetry
arguments and the superconvergence theorem on the generalized K-K relations, and comparing the results with the asymptotic behavior for large
values of $\omega_0$, new sum rules are obtained. We derive that all even moments of the real part and all odd moments of the imaginary parts
are null, except for the highest converging odd moment of the imaginary part of the susceptibility, which is directly related to the short time
behavior of the system. Furthermore, these results are also extended to the powers  $\left[\chi^{(n)}(\omega_0,\ldots,\omega_0)\right]^m$,
$m\geq 1$ of the susceptibility, and additional constraints are derived. The obtained generalized K-K relations and sum rules can be used to
check any experimental data and approximate theory of nonlinear phenomena, because they are necessary constraints which have to be obeyed. These
results generalize and extend what obtained by Ruelle \cite{Rue98,Rue98b} for Axiom A systems, set in a much more general theoretical framework
previous findings obtained for near equilibrium optical processes \cite{Luc03,Luc05} and  and simple yet prototypical mechanical systems near
equilibrium \cite{Luc98}, and shed light on the generality of the constraints deriving from the principle of causality, which can be used for
testing model outputs and experimental data, both for equilibrium and non-equilibrium systems. Note that, as discussed in
\cite{Peip99,Luc03,Luc05}, basically all K-K relations and sum rules can be rephrased, after lengthy but straightforward calculations, in terms
of absolute value and phase of the susceptibility function, which in some cases may be of easier experimental observation.

It is somewhat surprising, and encouraging in the perspective of the theory here developed, to see that the linear susceptibility of the Lorenz
system investigated in \cite{Reick02} looks a lot like the result of an optical experiment: the peaks of the imaginary part, corresponding to
the resonances of system (even if in this case there are no deterministic \textit{natural frequencies}) match the dispersive structures found
for the real part of the susceptibility. Note that, far from being a curiosity, it is through this approach that the optical constants of most
solids have been actually computed \cite{Bas83,Asp85}.

In order to clarify and complete the picture, we have shown, in App. \ref{kubo}, that the functions derived for non equilibrium steady states
are formally equivalent, at all perturbative orders, to what obtained with the Kubo formalism for the response of systems close to equilibrium,
apart from the measure involved in the phase space integration. In the case of near to equilibrium system, the measure is the one describing the
canonical distribution, whereas in the setting analyzed by Ruelle, the SRB measure of the unperturbed flow is involved. Therefore, all the
results presented in the paper apply, \textit{a fortiori}, for these equilibrium systems.

The response theory for Axiom A systems can have interesting implications for climate studies. In fact, the possibility of defining a response
function basically poses the problem of climate change is well-defined context, and, when considering the procedures aimed at improving climate
models, justifies rigorously the procedures of tuning and adjusting of the free parameters. Furthermore, qualitative differences between
different and widespread \textit{ensemble simulation} practices can be interpreted in this context. Moreover, the non-equivalence of free and
forced fluctuations explains why many attempts of applying the fluctuation-dissipation theorem in climate studies have basically failed.
Instead, it may be that the general theory of Kramers-Kronig relations described in this paper, which, in the case of non-equilibrium system, is
decoupled form the fluctuation-dissipation theorem, may provide a viable way of defining a comprehensive self-consistent theory of climate
change, ensured by the integral relations connecting the in-phase and our of phase components of the response of the system to external
perturbations. This is discussed in some greater detail in App. \ref{climate}.

We conclude with some practical caveats. As well known, it is surely not trivial in practical terms to effectively verify the K-K relations and
sum rules on experimental or model generated data. One general problem is their integral formulation, which requires that data are available on
a rather extensive spectral range and with a reasonable resolution. This may raise issue of computational costs and/or experimental set-up. The
extrapolations in K-K analysis can be a serious source of errors \cite{Asp85,Pei91}. Recently, King \cite{King02} presented an efficient
numerical approach to the evaluation of K-K relations, and singly and multiply subtractive K-K relations have been proposed in order to relax
the limitations caused by finite-range data \cite{Pal98,Luc03c}. It should be noted that K-K relations for higher-order susceptibilities are,
somewhat counter-intuitively, sometimes easier to verify than the linear K-K relations, because they have typically a much faster asymptotic
decrease. Whereas we have shown that at all orders large families of K-K relations hold for the various moments and various powers of the
susceptibility functions, it should be expected that they do not converge at the same rate when data of finite precisions coming from a finite
spectral range are used. See the discussion in \cite{Luc03,Luc05}. Furthermore, when considering chaotic systems, further problems in signal
detection of the system response at specific frequencies are related to the presence of a continuous spectrum in the background; this latter
issue may become more serious when nonlinear processes are examined and the observed monochromatic signal is weaker. Nevertheless, along the
line of Reick \cite{Reick02} these problems may result to be manageable.

\acknowledgments The author wishes to thank F. Bassani, K.-E. Peiponen, D. Ruelle, and A. Speranza for crucial intellectual stimulations.

\clearpage
\newpage
\appendix
\section{Reconciling Kubo's and Ruelle's general perturbative response functions}
\label{kubo}

In this appendix we show how to reconcile formally the $n^{th}$ order perturbative response for general systems characterized by non-equilibrium
steady state presented in Eq. \eqref{greenrue} with the classical results obtained with the Kubo formalism \cite{Kubo57} for the response for
systems close to equilibrium. Therefore, all the results presented in the paper apply, \textit{a fortiori}, for these equilibrium systems.

We consider a system of N degrees of freedom described by the canonical coordinates $q=(q_1,\ldots,q_N)$ and $p=(p_1,\ldots,p_N)$ and evolving
under the action of the Hamiltonian operator $H(q,p)=H_0(q,p)+h(q,p,t)$, composed of the unperturbed Hamiltonian $H_0(q,p)$ plus the time
dependent perturbation (weak) Hamiltonian expressed in the form $h(q,p,t)=-e(t)B(q,p)$ \cite{Zubarev}. The evolution equation of the system can
then be written as:
\begin{equation}
\dot{x}=F(x)+e(t)X(x)
\end{equation}
where $x=(q,p)$; $F(x)=\Omega\nabla H(x)$, $X(x)=-\Omega\nabla B(x)$, with $\Omega$ indicating the symplectic matrix. We assume that, if the
perturbation is set to 0, the expectation value of any observable $\Phi$ can be expressed as the following:
\begin{equation}
\langle \Phi \rangle_0 = \int \textrm{d}x \rho_0(x)\Phi(x)=\int \rho_0(\textrm{d}x)\Phi(x)
\end{equation}
where integration is performed in the phase space of the system, and the canonical distribution, which is absolutely continuous with respect to
the Lebesgue measure of the phase space, is defined as usual as:
\begin{equation}
\rho_0(\textrm{d}x)=\rho_0(x)\textrm{d}x=\frac{\exp{[-H_0(x)/kT]}}{\int \textrm{d}\Gamma
\exp{[-H_0(x)/kT]}}\textrm{d}x=\frac{\exp{[-H_0(x)/kT]}}{\int \rho_0(\textrm{d}x)}\textrm{d}x.
\end{equation}
Following the perturbative approach introduced by Kubo \cite{Kubo57}, we have that, for small perturbations, the expectation value of $\Phi$ at
time $t$ can be written as:
\begin{equation}
\langle \Phi \rangle(t) = \langle \Phi \rangle_0 + \sum_{n=1}^{\infty} \langle \Phi \rangle^{(n)}(t)
\end{equation}
where the terms under summation describe the non-equilibrium properties - for a system which is close to equilibrium - at all orders of
perturbation; in particular the $n=1$ terms provides information of the linear response of the system. The perturbative terms can be expressed
as follows \cite{Kubo57,Zubarev}:
\begin{eqnarray}
&&\langle \Phi \rangle^{(n)}(t) =\int\limits_{-\infty}^\infty\int\limits_{-\infty}^{\infty}\ldots
\int\limits_{-\infty}^{\infty}\textrm{d}\sigma_1\textrm{d}\sigma_2\ldots
\textrm{d}\sigma_n\nonumber\times\\
&&\times \Theta(\sigma_1)\Theta(\sigma_2-\sigma_1)\ldots\Theta(\sigma_n-\sigma_{n-1}) f(t-\sigma_1)f(t-\sigma_2)\ldots f(t-\sigma_n)\times \nonumber\\
&&\times \langle \left[B(x),\ldots \left[B(x(\sigma_n-\sigma_2)),\ldots
\left[B(x(\sigma_n-\sigma_1)),\Phi(x(\sigma_n))\right]\right]\ldots\right] \rangle_0=\nonumber\\
&&=\int\limits_{-\infty}^\infty\int\limits_{-\infty}^{\infty}\ldots \int\limits_{-\infty}^{\infty}\textrm{d}\sigma_1\textrm{d}\sigma_2\ldots
\textrm{d}\sigma_n G^{(n)}(\sigma_1,\ldots,\sigma_n)e(t-\sigma_1)e(t-\sigma_2)\ldots e(t-\sigma_n),
\end{eqnarray}
where the time evolution of the observables $B$ and $\Phi$ is due to the unperturbed Hamiltonian $H_0$. The two main are that the
non-equilibrium response is expressed as a convolution integral where the Kernel, which is the $n^{th}$ order Green function
$G^{(n)}(\sigma_1,\ldots,\sigma_n)$ is causal.

Note that the Kernel operator in the quantum case, where we deal with N-particles Hilbert space and observables are replaced by operators, is
formally obtained by simply substituting each Poisson brackets $[\bullet,\bullet]$ with $1/(\textrm{i}\hbar)$ times the commutator
$\{\bullet,\bullet\}$, and by redefining the expectation value at equilibrium of a generic operator $P$ as follows:
\begin{equation}
\langle P \rangle _0= \frac{\sum_a \langle a|P|a\rangle \exp{[-E_a/kT]}}{\sum_b \exp{[-E_b/kT] }}.
\end{equation}
where $|a\rangle$ is the eigenstate with eigenvalue $E_a$ of the Hamiltonian operator $H_0$.

Since the following trivial identity holds:
\begin{equation}
\left[B(x),\bullet \right]=X(x)\nabla(\bullet)=\Lambda(\bullet)
\end{equation}
and since, by definition, the evolution of any observable $A$ driven by the unperturbed Hamiltonian $H_0$ can be formally represented as
follows:
\begin{equation}
A(x(\tau))=\exp(i\tau L)A(x)=\Pi(\tau)A(x),
\end{equation}
where $iLA(x)=[A(x),H_0(x)]$, the $n^{th}$ order Green function can be formally written in the following compact and form:
\begin{equation}
G^{(n)}(\sigma_1,\ldots,\sigma_n)=\int
\rho_0(\textrm{d}x)\Lambda\Pi(\sigma_n-\sigma_{n-1})\ldots\Lambda\Pi(\sigma_2-\sigma_1)\Lambda\Pi(\sigma_1)\Phi(x).
\end{equation}
which is fully equivalent to the formula shown in Eq. \eqref{greenrue}, provided that the measure describing the equilibrium canonical
distribution is substituted with the general SRB measure.

\clearpage
\section{Response theory for non-equilibrium steady states and climate research}
\label{climate}

When adopting the chaotic hypothesis, the possibility of defining a response function of a perturbed non-equilibrium steady state and its actual
properties seem to have very interesting impacts in climate studies. On one side, this creates a context where the problem of climate change is
well-posed at mathematical level and where, when considering the procedures aimed at improving climate models, the tuning and adjustment of the
free parameters - at least locally - may be considered as a well-defined operation devoid of catastrophic impacts on the statistical properties
of the system. On the other hand, straightforward applications of fluctuation-dissipation theorem \cite{Leith75,Lin84}, or the idea that climate
change signals project on the natural modes of climate variability \cite{Corti99} seem inadequate, as discussed in \cite{Luc07}. Instead, it
seems that the theory of Kramers-Kronig relations described in this paper may provide a viable way of defining a comprehensive self-consistent
theory of climate change, ensured by the integral relations connecting the in-phase and our of phase components of the response of the system to
external perturbations. As an example, we may interpret Eq. \eqref{sr2} as the fact that the static response function - measuring
\textit{climate sensitivity} - can be related to the out-of-phase response to same forcing at all frequencies, at least in first approximation.

The concepts behind the Ruelle response theory also clarify the meaning of some common \textit{ensemble simulation} practices, which are widely
adopted by the climate modelling community with the goal of estimating the uncertainty on the statistical properties of the model outputs, when
a specific set of observables is considered \cite{Luc02,SPELUC,IPCC}. Three different strategies, which are nevertheless more and more
hybridized, can be pointed out:
\begin{itemize}
\item Each simulation is performed with the same climate model, but starting from slightly
different initial state;
\item Each simulation is performed with the same climate model, but with slightly different values of some key uncertain parameters characterizing
the global climatic properties,
\item Each simulation is performed with a different climate model (\textit{multi-model ensemble}).
\end{itemize}
Under the chaotic hypothesis, the first procedure seems useful, since a more detailed exploration of the phase space of the system, with a
better sampling - on a finite time - of the attractor of the model. The significance of the second procedure seem to be reinforced by the
response theory for non equilibrium steady states, because in this case the variously tuned models basically explore parameterically deformed
ergodic measures, and the macroscopic \textit{sensitivity} of the model is thus explored. As for the third procedure, whereas it surely allows
for climate model intercomparison, aggregating information from from rather different attractors seems ill-defined.

\end{document}